\documentclass[nonacm,sigplan]{acmart}

\usepackage[]{hyperref}
\usepackage{multirow}
\usepackage{xspace}

\newcommand{\costalg}{\texttt{Cost-GLB}\xspace}
\newcommand{\allalg}{\texttt{All-GLB}\xspace}
\newcommand{\eglb}{\texttt{E-GLB}\xspace}
\newcommand{\ouralg}{\texttt{SE-GLB}\xspace}

\begin{document}

\title{Towards Socially and Environmentally \\
Responsible AI}

 \author{Pengfei Li}
 \authornote{Equal contribution}
 \affiliation{%
   \institution{UC Riverside}
   \country{}
   \city{}
   }

 \author{Yejia Liu}
 \authornotemark[1]
 \affiliation{%
   \institution{UC Riverside}
   \country{}
   \city{}
   }

 \author{Jianyi Yang}
 \affiliation{%
   \institution{UC Riverside}
   \country{}
   \city{}
   }

 \author{Shaolei Ren}
 \authornote{Corresponding author: Shaolei Ren (shaolei@ucr.edu)}
  \affiliation{%
   \institution{UC Riverside}
   \country{}
   \city{}
   }

\begin{abstract}

The sharply increasing sizes of  artificial intelligence (AI) models
come with significant energy consumption
and environmental footprints, which can disproportionately impact certain (often marginalized) regions and hence create environmental inequity concerns. Moreover, concerns with social inequity have also emerged, as AI computing resources
may not be equitably distributed across the globe and
users from certain disadvantaged regions with severe resource constraints 
can consistently experience inferior model performance.
Importantly, the inequity concerns that encompass both social and environmental dimensions still remain unexplored and have increasingly hindered responsible AI. In this paper, we 
 leverage the spatial flexibility of AI inference workloads
 and propose equitable geographical load balancing
 (GLB) to fairly
 balance AI's regional social and environmental costs. Concretely, to penalize the disproportionately high
 social and environmental costs for equity,
 we introduce $L_q$ norms as novel regularization terms into the optimization objective
 for GLB decisions.
Our empirical results based on real-world AI inference traces
demonstrate that while the existing GLB algorithms result
in disproportionately  large 
social and environmental costs in certain regions, 
our proposed equitable GLB can fairly balance AI's negative social and environmental costs across all the regions. 
\end{abstract}

\maketitle 
\pagestyle{plain} 

\section{Introduction}

In the rapidly evolving field of artificial intelligence (AI), a significant transformation is underway with the emergence of large
foundation models as exemplified by Large Language Models (LLMs) like GPTs~\cite{brown2020language} and Vision Transformers Models (ViTs)~\cite{dosovitskiy2021image}. These cutting-edge AI models demonstrate the ability to function effectively in diverse contexts, engaging with extensive vocabularies and image data for unforeseen AI tasks, i.e., zero-shot abilities. 
To serve inference requests, they are typically deployed across geographically distributed data centers for better
service availability, lower transmission latency, and/or privacy regulations.

\textbf{Environmental inequity.} Powerful yet hungry, large AI models require substantial resources not only during training but also in deployment and inference. For some popular AI services
such as text and image generation, the total energy consumption for inference can be comparable
to or even exceed that for training, resulting in
huge carbon emissions and freshwater usage 
\cite{Carbon_SustainbleAI_CaroleWu_MLSys_2022_wu2022sustainable,Shaolei_Water_AI_Thirsty_arXiv_2023_li2023making}.
To curb the growing environment footprint, many recent efforts have been devoted to enhancing the efficiency and reducing the energy consumption of AI models. Example strategies include model compression that reduces AI's computational demand for inference (typically at a sacrifice of model performance)~\cite{lin2023awq, ma2024llm} and  geographical load balancing (GLB) 
that leverages spatial heterogeneities to route more workloads to low-cost and/or greener regions~\cite{li2023environmentally,GLB_AI_Inference_NoLatencyImpact_AndrewChien_Chicago_HotCarbon_2023_10.1145/3604930.3605705}. Additionally, on the infrastructure side, there has been a rise in the adoption of carbon-free energy
and climate-conscious cooling system designs in the data center industry. For instance, utilizing air-side economizers where climate conditions allow has become increasingly common to cut the direct water consumption~\cite{meta2023water}. 

While these approaches can effectively minimize AI's total environmental footprint, the rise of
\emph{environmental inequity} --- AI's negative environmental impact
can disproportionately affects certain (often marginalized) regions 
\cite{Justice_AINowInstitute_CallforEnvironmentJusticeAI_2021, li2023environmentally} --- has become increasingly worrisome, potentially leading to other unintended social and ecological consequences and widening regional disparities. Importantly, the disproportional distribution of
AI's environmental cost across different regions can be amplified
by existing approaches to managing AI systems (e.g., load distribution and AI model scaling) that often prioritize
the total environmental cost rather than the cost borne by individual
 regions which are most environmentally vulnerable \cite{li2023environmentally}. Compounded
by the sharply growing demand, AI's environmental inequity has
received calls for mitigation efforts from various organizations,
such as UNESCO \cite{Justice_Policy_Ethical_AI_Recommendation_UNESCO_2022},
Meta \cite{meta2021sustainability} and the State of California \cite{Justice_AI_EnvironmentalEquity_CaliforniaReport_2023}.

\textbf{Social inequity.}  Going beyond environmental footprints, 
concerns with AI's social inequity have also emerged \cite{unitednations2024}.
For now, only a few major tech players have the resource and capacity to train and deploy large AI models. Thus,
due to the uneven deployment of computing resources across
the globe, users from different regions may encounter varying AI model sizes
and performances (e.g., larger AI models typically imply better inference performance in terms of the accuracy and task scores), leading to complex societal consequences.
For example, studies have indicated that people are becoming increasingly reliant on LLMs for acquiring knowledge, suggesting that subpar LLMs could jeopardize the prospects of these individuals~\cite{Tokayev2023}. Thus,
AI's potentially unfair model performance has close relevance to its social inequity.
Crucially, the existing environmentally conscious approaches to AI system management
(e.g., choosing larger AI models with better performances/accuracies when
there are more solar energy available) may further reinforce AI's performance unfairness among users from different regions, enlarging the social inequity.

\textbf{Contributions.}
With the growing need for AI as a public resource serving the broader society, it becomes increasingly imperative to rectify AI's emerging social and environmental inequities and enable truly responsible AI~\cite{rayhan2023ethical, meta2021sustainability}.  In this paper, we focus on the AI inference stage and introduce a novel equity-aware
GLB algorithm to fairly balance
AI's social and environmental costs across different regions. More specifically, we consider the performance cost of heterogeneous AI models and the carbon and water footprints associated with AI model inference by dynamically scheduling users' inference requests (a.k.a. workloads) using GLB. 
When optimizing GLB decisions, we leverage $L_q$ norms
in terms of AI's social and environmental costs as regularization terms to penalize decisions that disproportionately affect certain regions. In other words, regions with higher environmental and/or social
costs will be prioritized and given a larger weight when leveraging GLB to minimize the total cost. By doing so, both the social and environmental costs of AI inference are more evenly distributed across different regions, thus mitigating AI's social and environmental inequities.

To assess the effectiveness of our method on promoting socially
and environmentally equitable AI, we conduct a simulation-based case study of $10$  geographically-distributed data centers serving inference requests
for an LLM over an 18-day period.
Our empirical results demonstrate that 
while the existing GLB algorithms result
in disproportionately  large 
social and/or environmental costs in certain regions, 
our proposed equitable GLB can fairly balance AI's negative social and environmental costs across all the regions.

\section{Related Works}

From the social fairness perspective, much attention has been directed towards protecting groups with certain attributes~\cite{wan2023processing,  pessach2022review, li2023fairer}. The issue is partially rooted in inherent biases within datasets and could potentially be exacerbated by models~\cite{li2023fairer, wan2023processing}.
To address such unfairness, numerous strategies have been developed. For instance, ~\cite{Biswas2021, madras2018learning} suggest removing sensitive attributes from datasets to prevent the model from relying on them, while others adjust prediction outcomes after training ~\cite{pessach2022review, noriegacampero2018active}. 
Additionally, some have advocated for equivalent metrics, such as error rates, among specific groups~\cite{cotter2019optimization,baharlouei2020renyi}. These studies
typically focus on the model training stage, but the attained fairness
can be compromised if AI models of different sizes are not equitably chosen for users from different regions.
By stark contrast, we focus on the AI inference stage and judiciously
balance the user requests from different regions across geographically distributed data centers hosting heterogenous AI models.

To address AI's environmental impacts,  existing studies primarily focus on minimizing environmental metrics such as the total carbon emission, water footprint, or a weighted combination thereof, to enable environmentally responsible AI model training
and inference \cite{GLB_AI_Inference_NoLatencyImpact_AndrewChien_Chicago_HotCarbon_2023_10.1145/3604930.3605705,Shaolei_Water_AI_Thirsty_arXiv_2023_li2023making, Carbon_SustainbleAI_CaroleWu_MLSys_2022_wu2022sustainable}. Nonetheless, concerns with AI's environmental inequity across different regions have remained largely unaddressed.
A recent study ~\cite{li2023environmentally} has proposed to tackle 
the uneven distribution of AI's regional 
environmental costs via GLB.
But, this approach overlooks the social equity dimension, which is equally, if not more, important element of responsible AI.

\section{Problem Formulation} 

We focus on the AI inference stage 
and consider a set of pre-trained AI models denoted by $\mathcal{K} = \{1,2,\cdots, K\}$, each with different performance and energy consumption for serving an inference request. There
are a set of geographically distributed
data centers $\mathcal{N} = \{1,2,\cdots, N\}$ serving
users coming from a set of regions
$\mathcal{J} = \{1,2,\cdots, J\}$.

\textbf{Operational cost.}
At each time $t$,  data center $i$ dynamically selects one or more of the available heterogeneous AI models to serve the incoming workloads. More formally, we denote $y_{i,j}^k(t) \geq 0$ as the workload dispatched from region $j$ to data center $i$ served through model $k$ at time $t$.
Given the scheduled demand $y_{i,j}^k(t)$,
we denote the energy consumption and computational resources necessary for deploying model $k$ in data center $i$ as $e_{i,k}(y^{k}_{i,j}(t))$ and $r_{i,k}(y^k_{i,j}(t))$, respectively.
For example, both $e_{i,k}(y^{k}_{i,j}(t))$ and $r_{i,k}(y^k_{i,j}(t))$ can be modeled as linearly increasing functions
in terms of $y^k_{i,j}(t)$.
Thus, the total energy consumption at data center $i$ can then be calculated as
\begin{equation*}
    e_i(t) = \sum_{j\in \mathcal{J}} \sum_{k \in \mathcal{K} } e_{i,k}(y^{k}_{i,j}(t)).
\end{equation*}

For notational simplicity, we define the set of workload distribution decisions at time $t$ as $y(t) = \{ y_{i,j}^k(t) | i\in \mathcal{N}, j\in \mathcal{J}, k \in \mathcal{K} \}$. We also take the energy price $p_{i,t}$ and power usage effectiveness (PUE, which accounts for non-IT energy overheads) $\gamma_i$ of data center $i$ into consideration. As a result, the total operational cost 
at time $t$ can be written as 
\begin{equation}
    cost_t(y(t))  =  \sum_{i\in\mathcal{N}}\gamma_i \cdot p_{i,t}\cdot\left[\sum_{j\in \mathcal{J}} \sum_{k \in \mathcal{K} } e_{i,k}(y^{k}_{i,j}(t))\right].
\end{equation}

\textbf{Social inequity cost.}
We define the noramlized performance cost of the AI model $k$ as $s_k(y_{i,j}^k(t))=s_k\cdot y_{i,j}^k(t) \geq 0$, where $s_k\geq0$ represents the inference performance degradation cost for each request when using model $k$ compared to the best possible model (usually the largest model~\cite{llama_2_touvron2023llama}). For example, when model $l$ has the best performance, its performance cost is zero for any allocated request. Here,
the performance cost can be measured
in terms of various metrics of an AI model (e.g., average inference accuracy and score of an LLM for a set of target tasks, among others).
Thus, the total performance cost of the workload from region $j$ is computed as $\sum_{i \in \mathcal{N}} \sum_{k \in \mathcal{K} } s_k(y_{i,j}^k(t))$, which, when normalized by the total workload $\lambda_{j,t}$, represents the AI model's  
average social performance for users from region $j$ (i.e., a type of \emph{group} fairness \cite{pessach2022review}). To balance AI's performance for users from different regions, we introduce a social fairness function $f_t(y(t))$ 
in terms of $L_q$ norm of the average performance costs for users from different regions:
\begin{equation}
    f_t(y(t)) = \left[ \sum_{j \in \mathcal{J}} \left[ \frac{\sum_{i \in \mathcal{N}} \sum_{k \in \mathcal{K} } s_k(y_{i,j}^k(t))}{\lambda_{j,t}} \right]^q \right]^{1/q},
\end{equation}
where $q\geq1$ is a hyperparameter that promotes AI's social equity for users from different regions. Concretely, we only care about
the average AI model performance across different regions when $q=1$ (i.e., no consideration of AI's social equity),
whereas we focus on minimizing
AI's worst regional model performance when $q\to\infty$ (i.e., solely considering AI model performance for users from the most disadvantaged regions). The priorities for these two conflicting objectives are adjusted by varying $q\geq1$.

\textbf{Environmental inequity cost.}
Carbon emissions associated with fossil fuels and water consumption are the two main non-negligible factors. 
Besides the global warming effects, carbon emissions have significant local effects such as high air pollution and even elevated immortality rates \cite{Carbon_LocalImpact_Stanford_2010_doi:10.1021/es903018m}, thus making it necessary to balance AI's regional carbon emissions.
Depending on the fuel mix for electricity generation, the carbon emission rate can vary significantly across different physical locations and times of the day.
Specifically, the carbon emission of data center $i$ is denoted as $c_{i,t} (e_{i}(t))$, where $e_{i}(t)$ is the total energy consumption for running AI inference in data center $i$ at time $t$. 
In general, an increased proportion of carbon-intensive energy sources (e.g. hard coals) directly correlates with higher carbon emissions, impacting the function $c_{i,t}(\cdot)$. 
The water consumption of deploying AI models is another important environmental cost and can be divided into two categories: onsite and offsite \cite{Shaolei_Water_AI_Thirsty_arXiv_2023_li2023making}. 
For each data center, onsite water is
evaporated to  
reject the heat generated by servers
into the outside environment (if the data center uses cooling towers), or cool and humidify the air entering the data center (if the data center uses air-side free cooling) \cite{Shaolei_Water_AI_Thirsty_arXiv_2023_li2023making}.
The offset water refers to the water consumed for the electricity generation. In total, we define the water consumption as $w_{i,t} (e_{i}(t))$, which considers both onsite and offsite water and is linearly increasing with $e_{i}(t)$ depending on
the runtime water usage effectiveness.

The total environmental cost of data center $i$ is defined as 
\begin{equation*}
    \mathcal{H}_{i}(\sum_{t=1}^T y(t)) = \sum_{t=1}^T \biggl[ \mu_c  c_{i,t}\bigl(e_i(t)\bigr) + \mu_w  w_{i,t}\bigl( e_i(t) \bigr) \biggr],
\end{equation*}
where the hyperparameters $\mu_w\geq0$ and $\mu_c\geq0$ convert the carbon emission and water consumption to a single unit cost and balance their relative importance. 
By applying the $L_q$ norm,
the overall environmental inequity cost is defined as
\begin{equation}\label{eqn:environmental_inequity_cost}
    g(\sum_{t=1}^T y(t)) = \left[ \sum_{i\in \mathcal{N} } \left( \mathcal{H}_{i}(\sum_{t=1}^T y(t)) \right)^q \right]^{\frac{1}{q}},
\end{equation}
where $q\geq1$ prioritizes the minimization of AI's environmental cost in more disadvantaged data center locations/regions. In particular,
when $q\to\infty$, \eqref{eqn:environmental_inequity_cost} becomes AI's worst environmental impact over all the data center locations.

\textbf{GLB objective.} We formulate the optimization objective of our socially and environmentally equitable GLB (called \ouralg)  as follows:
\begin{subequations}\label{eqn:objective_offline_heterogeneous}
 \begin{gather}\label{eqn:objective_heterogeneous}
 \begin{gathered}
        \min_{y(t),t=1,\cdots,T} \sum_{t=1}^T {cost}_t(y(t)) + \sum_{t=1}^T f_t(y(t)) + g \left(   \sum_{t=1}^T y(t)  \right)\\
        +\sum_{t=1}^T\sum_{i\in\mathcal{N},j\in\mathcal{J},k\in\mathcal{K}}y_{i,j}^k(t)\cdot d_{ij},
\end{gathered}
       \\
       \label{eqn:constraint_gateway_heterogeneous}
 s.t. \;\;\;\;\;\;        \sum_{i\in\mathcal{N}} \sum_{k\in\mathcal{K}}y_{i,j}^k (t) = \lambda_{j,t}, \forall
       \; j\in\mathcal{J}, t=1,\cdots,T,\\
\label{eqn:constraint_datacenter_capacity}
      \;\;\;\;\;\;   \sum_{k\in\mathcal{K}}r_{i,k}\left(\sum_{j \in \mathcal{J}} 
 y_{i,j}^k(t)\right)\leq M_i,  \forall
       \; i\in\mathcal{N},t=1,\cdots,T
 \end{gather}
\end{subequations}
In \eqref{eqn:objective_heterogeneous},
the term $\sum_{i\in\mathcal{N},j\in\mathcal{J},k\in\mathcal{K}}y_{i,j}^k(t)\cdot d_{ij}$ accounts for the total moving cost for scheduling user requests from region $j$ to data center $i$, where $d_{ij}$ represents the  moving cost for scheduling one unit of request (e.g., in proportion to the distance between region $j$ and data center $i$). The constraint \eqref{eqn:constraint_gateway_heterogeneous} means that we need to schedule all the user demand $\lambda_{j,t}$
for each region $j$ without request dropping, and the constraint \eqref{eqn:constraint_datacenter_capacity} denotes the computational resource constraint for AI inference in each data center $i$. Note that we can also easily add other constraints such as workload routing constraints
(i.e., user requests from
region $j$ can only be routed to certain data center locations due to data sovereignty regulations or latency constraints).

Compared to the existing literature on GLB that typically minimizes the total cost or focuses on the environmental impact \cite{gao2012s,li2023environmentally},
the key novelty of our formulation is
to holistically address AI's social and environmental inequities by using
$L_q$ norms to penalize GLB decisions that lead to
disproportionately high social and/or environmental costs in certain disadvantaged regions.

\section{A Case Study}

We run a simulation study to preliminarily validate \ouralg
to mitigate AI's social and environmental inequities.

\subsection{Setup}
We consider 10 geographically distributed data centers: four in the U.S. (Virginia, Georgia, Texas, and Nevada), four in Europe (Belgium, the Netherlands, Germany, and Denmark), and two in Asia (Singapore and Japan). Each of these locations hosts a large number of data centers. We also consider 10 regions, each corresponding to one distinct data center location in our experiments. 
To highlight the potential of equity-aware GLB, we consider \emph{full} GLB flexibility, where workloads can be dispatched from any region to any data center. To host an LLM inference service,
each data center contains a cluster of 150 identical Nvidia DGX A100 servers each equipped with eight NVIDIA A100 GPUs and a maximum power of $6.5$kW. Excluding other services beyond our scope, each data center has a maximum AI inference server power of  $~\sim{1}$ MW. The data center PUE is set as $1.1$ to adhere to efficient operation standards. The regional environmental impact is assessed using a weighted combination of carbon and water footprints. For inference, we assume three LLMs of different sizes are available: Llama-2-7B, Llama-2-13B, and Llama-2-70B~\cite{llama_2_touvron2023llama}.

\textbf{Datasets.} We utilize the GPU power trace spanning 18 days as used in \cite{li2023environmentally}.
We gather evaluation scores of Llama-2 from HuggingFace~\cite{wolf2020huggingfaces} across the model sizes of $7B$, $13B$, and $70B$ on benchmarks AI2 Reasoning Challenge, HellaSwag,
and Truthful QA. We then average and normalize these scores for measuring AI's performance costs. Hourly energy prices across the 10 data centers are obtained from~\cite{iea_data_statistics} for Europe and Asia, and from their respective ISOs for U.S. data centers ~\cite{us_eia_opendata}. Hourly weather data from~\cite{iowa_mesonet} is utilized to calculate wet bulb temperature from dry bulb temperature and relative humidity. On-site WUE is determined using an empirical formula from \cite{water_saving_islam2016exploiting}.

\textbf{Evaluation metrics.} We consider four metrics: 1) \emph{average energy cost}, calculated as the total energy cost over 18 days divided by the number of data center locations; 2) \emph{average environmental footprint and social cost}, representing the total carbon emission, water footprint, and performance cost by the number of data center locations; 3) \emph{maximum regional environmental footprint and performance cost}, which identifies the highest environmental and performance costs among the 10 data center locations and user regions; 4) \emph{{max/avg} ratio}, representing the ratio of the maximum cost to the average cost for relative comparison. A lower value on this metric indicates a more equitable solution.

\begin{table}[!t]
   \scriptsize
  \caption{Comparison between different GLB algorithms.}   \label{table:long_sequence}
  \centering
  \begin{tabular}{c |c|c|c|c|c} 
  \toprule
   \multicolumn{2}{c|}{\multirow{2}{*}{\textbf{Metric}}} & \multicolumn{4}{c}{\textbf{Algorithm}} \\ 
  \cline{3-6}
  \multicolumn{2}{c|}{}  &  \costalg &  \allalg  & \eglb   & \textbf{\ouralg} \\ 
  \hline
  \textbf{Energy} (US\$)  & avg  & 83524 & 92945 & 101106 & 108197 \\ 
  \cline{1-6}
  \multirow{3}{*}{\textbf{Water} ($\text{m}^3$)} & avg   & 476.72 & 465.44 & 433.24 & 456.58\\ 
  \cline{2-6}
  & max   & 1410.92 & 842.44 & 652.72 & 649.41  \\
  \cline{2-6}
  & \textbf{max/avg} & 2.96 & 1.81 & 1.51 & 1.42 \\
  \cline{1-6}
  \multirow{3}{*}{\textbf{Carbon} (ton)}   & avg  & 36.720 & 32.090 & 29.548 & 32.163 \\ 
  \cline{2-6}
  & max  & 110.275 & 55.491 & 41.923 & 48.054 \\ 
  \cline{2-6}
  &  \textbf{max/avg} & 3.00 & 1.73 & 1.42 & 1.49  \\
  \cline{1-6}
  \textbf{Normalized}   & avg  & 0.262 & 0.244 & 0.268 & 0.248   \\ 
  \cline{2-6}
  \textbf{Performance} & max  & 0.449 & 0.353 & 0.313 & 0.253  \\  
  \cline{2-6}
  \textbf{Cost} &  \textbf{max/avg} & 1.71 & 1.45 & 1.17 & 1.02 \\
  \cline{1-6}
  \textbf{Performance}   & avg  & 57.83 &  58.11 & 57.74 & 58.05   \\ 
  \cline{2-6}
  \textbf{Score} & min  & 54.94 & 56.43 & 57.04 & 57.98 \\ 
  \bottomrule
  \end{tabular}
\end{table}

\textbf{Baselines.} 1)\xspace\costalg: This algorithm optimizes the average energy cost and the performance cost. It can also be seen as a special case of \ouralg where $\mu_c$ and $\mu_w$ are set as zero and $q=1$. 
2)\xspace\allalg: This algorithm minimizes the weighted sum of the energy cost, environmental cost and societal cost (i.e., $q=1$) based on  \cite{water_saving_islam2016exploiting}. 
3)\xspace\eglb: The environmentally equitable GLB algorithm which is studied in 
 \cite{li2023environmentally} and does not address AI's social inequity.
Note that we do not consider the baseline that solely minimizes
energy cost, as this approach would simply force the data centers to always choose the smallest model for inference.

\subsection{Results}
We run an offline optimizer with all the future information in our case study, while online algorithms that optimize GLB without knowing future information are left as our future work.
In Table~\ref{table:long_sequence}, we show the cost comparison between baselines and \ouralg. By default, the weights assigned to carbon emission and water consumption are $\mu_w = 60$ (US\$/$m^3$) and $\mu_c = 1500$ (US\$/ton), unless otherwise specified. We can observe that \costalg has the lowest energy cost compared to other GLB approaches since it prioritizes the energy cost minimization. However, it also leads to the highest average carbon emission and water consumption. Additionally, \costalg exhibits the highest max to average ratio in terms of social
and environmental equities. Therefore, solely optimizing for energy cost can overburden certain regions with excessive workloads and worsen the AI's inequity. By comparison, \allalg takes a weighted sum of energy cost and the environmental footprint, which reduces average water consumption and carbon emission. \eglb further reduces the max to average ratio of environmental footprint by minimizing the $L_q$ norm of AI's environmental impact across different locations. 
\ouralg explicitly considers the $L_q$ norms of both social and environmental costs, thereby achieving a more equitable distribution of AI's
model performance and environmental impact across different user regions and data center locations. While this comes at an increased energy cost
due to the conflict between equity
and energy cost minimization, we argue that
 the cost increase is acceptable in order to mitigate AI's inequity that would otherwise create unintended socio-ecological consequences as people increasingly rely on AI.

\section{Concluding Remarks}
In this work, we holistically consider AI's social and environmental equity
and propose novel equity-aware GLB
to balance AI's regional social and environmental costs towards responsible AI.
Our key novelty is to introduce
$L_q$ norms to penalize GLB decisions
that would otherwise lead to disproportionately high social and/or environmental costs in disadvantaged regions. Our
 empirical evaluation has shown the effectiveness of our proposed approach in improving both social and environmental equity by prioritizing
the most disadvantageous data centers and user regions.

\section*{Acknowledgement}
This work was supported in part
by the U.S. NSF under grants 
CNS-1910208, CNS-2007115, and CCF-2324941.


\bibliographystyle{plain}
\bibliography{main.bbl}
\end{document}